\documentclass[conference]{IEEEtran}
\IEEEoverridecommandlockouts
\usepackage{amsmath,amssymb,amsfonts}
\usepackage{algorithmic}
\usepackage{graphicx}
\usepackage{textcomp}
\usepackage{xcolor}
\def\BibTeX{{\rm B\kern-.05em{\sc i\kern-.025em b}\kern-.08em T\kern-.1667em\lower.7ex\hbox{E}\kern-.125emX}}
    
\usepackage{acronym}
\usepackage{siunitx}
\usepackage{cuted}
\usepackage{subfigure,epsfig,xcolor, pstricks}
\usepackage{todonotes}

\begin{document}

\title{Complementary Semi-Deterministic Clusters for Realistic Statistical Channel Models for Positioning}
%Toward realistic statistical channel models for positioning: Complementary Semi Deterministic Clusters

\author{
    Mohammad Alawieh\IEEEauthorrefmark{1}, 
    Ernst Eberlein\IEEEauthorrefmark{1}, 
    Stephan Jäckel\IEEEauthorrefmark{2}, 
    Norbert Franke\IEEEauthorrefmark{1}, 
    Birendra Ghimire\IEEEauthorrefmark{1}, 
    Tobias Feigl\IEEEauthorrefmark{1},\\
    George Yammine\IEEEauthorrefmark{1}
    and Christopher Mutschler\IEEEauthorrefmark{1}\\
    \and
    \IEEEauthorblockA{
        \IEEEauthorrefmark{1}~Fraunhofer Institute for Integrated Circuits IIS, N\"urnberg, Germany\\
        \texttt{\footnotesize\{firstname.lastname\}@iis.fraunhofer.de}
    }
    \and
    \IEEEauthorblockA{
        \IEEEauthorrefmark{2}~SJC Wireless, Berlin, Germany\\
    	\texttt{\footnotesize jaeckel@sjc-wireless.com}
    }
}

%%%%%%%%%%%%%%%%%% abstract %%%%%%%%%%%%%%%%%%%%%%%

\maketitle
\begin{abstract}
Positioning benefits from channel models that capture geometric effects and, in particular, from the signal properties of the first arriving path and the spatial consistency of the propagation condition of multiple links. The models that capture the physical effects observed in a realistic deployment scenario are essential for assessing the potential benefits of enhancements in positioning methods. Channel models based on ray-tracing simulations and statistical channel models, which are current state-of-the-art methods employed to evaluate performance of positioning in 3GPP systems, do not fully capture important aspects applicable to positioning. Hence, we propose an extension of existing statistical channel models with semi-deterministic clusters (SDCs). SDCs allow channels to be simulated using three types of clusters: fixed-, specular-, and random-clusters. Our results show that the proposed model aligns with measurements obtained in a real deployment scenario. Thus, our channel models can be used to develop advanced positioning solutions based on machine learning, which enable positioning with centimeter level accuracy in NLOS and multipath scenarios. 

% fixed, specular, random habe ich in der intro nicht gefunden?
% disktuieren wir diese 3 auch wirklich?
%We show that we can generate realistic channel measurements by comparing simulation with reality.

%Our fine-grain channel simulation helps to develop advanced positioning solutions that leverage machine learning, enabling centimeter-level positioning.
\end{abstract}


%%%%%%%%%%%%%%%%%% intro %%%%%%%%%%%%%%%%%%%%%%%
% Status
% - First draft written by ebl
% - Reference to hybrid model as described in TR38.901
% General 3GPP positioning evaluations with reference to TR38.857 and TR38.855 and how its progressing to include precise positioning and AI aspects (if we can talk down on cluster correlation from multiple BSs, we highlight this as an enabler for FP (fingerprint) evaluation approaches) . 

\section{Introduction}\label{sec:introduction}

%For instance, 5G aims to support a wide range of applications with different requirements in terms of accuracy, latency, availability, and power consumption.
% unverständlich
%However, for most of these applications, positioning evaluation models that do not account for positioning effects are either unsuitable for specific methods or may not be sufficient to provide credible results.
% vorschlag:
%For most of these applications, classic positioning methods, such as least squares optimization for trilateriation, are inappropriate for multipath situations~\cite{fingerprinting_paper_from_niitso}.
%However, even modern learning-based localization methods trained on simulated data fail as the propagation environment is not modeled accurately enough~\cite{transferlearning_paper_from_stahlke}.

\begin{figure}[t!]
\centering
    \includegraphics[width=0.8\columnwidth]{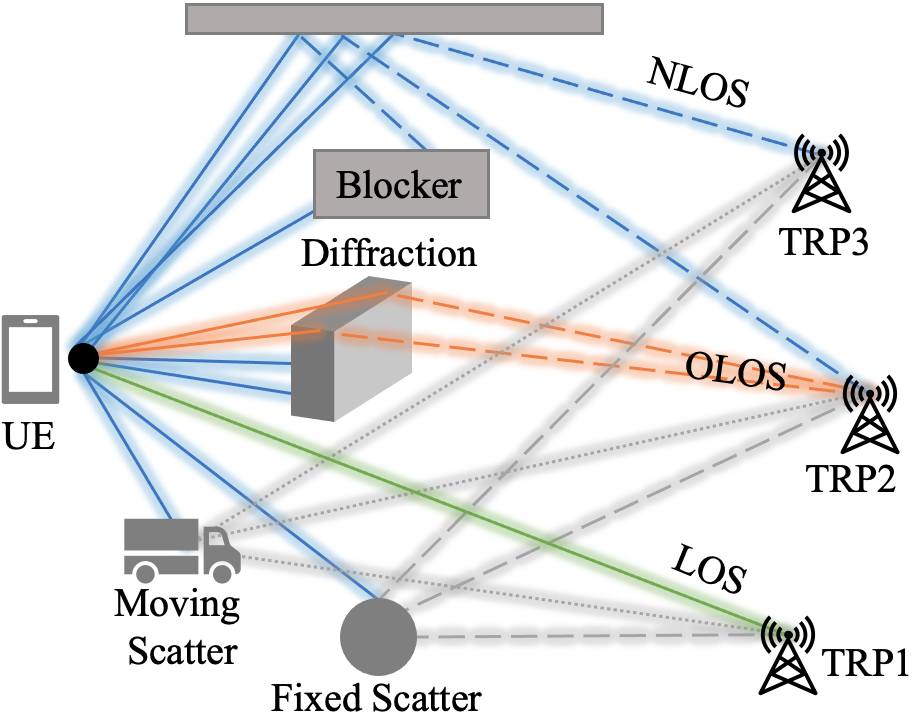}
    \caption{
    Overview of three realistic propagation scenarios: 
    (1) TRP1 is in the valid LOS state, i.e., the (green) LOS path is typically stronger than the multipath components (MPC), allowing for unconstrained positioning accuracies.
    (2) TRP2 is in OLOS state, i.e., the (orange) LOS path is blocked, but w.r.t. the size and nature of the obstacle, the direct path may still be recognizable with a slight delay; Diffraction effects may also generate low-delay MPCs relative to the direct path, so the OLOS state may already significantly affect positioning methods.
    (3) TRP3 is in NLOS state, i.e., the direct path is blocked, only multipath (blue) components are received that drastically affect classic positioning methods.}
    \label{fig:SDC_types}
\end{figure}

\noindent 
Location services in 3GPP have evolved to support a wide range of UE positioning applications, ranging from industrial scenarios to emergency calls. Support for demanding applications like controlling an automated guidance vehicle (AGV) in industrial environment require accuracy in centimeters region and latency in ms region~\cite{tr38855}.  
%Other applications like autonomous driving demands extremely high reliability of the position estimates~\cite{citation-missing}. %durch diesen satz entsteht ein Bruch; so ist es besser: aus dem Drohnen-Industrieszenario OLOS/NLOS erklären anhand der Fig. 1.
Fig.~\ref{fig:SDC_types} shows different propagation states in a positioning scenario for a UE in relation to 3 TRPs. Ideally, the simulation environment will reflect effects like the Obstructed Line-of-Sight (OLOS) and allow the usage of the fixed scatters despite the changes in the environment. The simulation environment needs to reflect these effects to enable a realistic evaluation of the positioning requirements.
 
Classical positioning methods, such as that relying on least squares optimization for trilateriation, are based on the assumption of Line-of-Sight (LOS) condition between the transmitter and receiver, do not perform well in Non-Line-of-Sight (NLOS) conditions~\cite{niitsoo2018convolutional}. Machine learning is a promising approach for achieving high accuracy positioning in such scenarios. However, data collection using actual measurements in the field is costly. Therefore, utilizing simulated data to train the model is one way to train the machine learning model~\cite{feigl2018RNN}. However, learning-based localization methods trained on simulated data fail if the propagation environment is not modeled accurately enough in the existing channel models.

%State-of-the-art positioning techniques employ machine learning. Therefore, they require spatially coherent information that represents the propagation environment to provide accurate positions even in challenging multipath environments~\cite{paper_ml4pos}. Walls, floors or other flat objects may create reflections.
% hier muss eine brücke zum eigentlich problem kommen ...
% wir kennen die probleme (fixed, specular, diffuse / random), die modelle berücksichtigen diese nicht, daher erweitern wir die modelle in unserem papier?
%For specular reflections, the angle of reflection may be equal to the angle of incidence.
%Thus, the effective reflection point may depend on the relationship between UE, TRP, and reflector position, as well as the orientation of the reflector.
%However, the reflected signal may also be radiated in different directions or vary due to moving scatterers.
%Hence, a potential statistical channel model must embed this spatially coherent information to enable machine learning to predict accurate positions. Fig.~\ref{fig:SDC_types} shows different propagation states in a positioning scenario for a UE in relation to 3 transmission and reception points (TRPs).

To account for the above identified issues and also maintain the generality of the statistical models, we extend the statistical models to include semi-deterministic clusters (SDC) and assess the impact of the SDCs on both channel properties and positioning accuracy using real reference measurements. Our experiments show that our proposed channel model extensions generate realistic measurements. Hence, we propose to use our channel model to develop an advanced positioning solution based on machine learning to enable realistic measurements which also cover OLOS situations.

The rest of this paper is structured as follows. 
Sec.~\ref{sec:relatedwork} reviews existing channel models. 
Sec.~\ref{sec:sdc_concept} introduces the concept of SDCs. 
Sec.~\ref{sec:setup} describes our measurement and simulation setups before Sec.~\ref{sec:evaluation} discusses the results. 
Sec.~\ref{sec:conclusion} concludes.

%%%%%%%%%%%%%%%%%% related work %%%%%%%%%%%%%%%%%%%%%%%
\section{Overview on TR38.901 positioning aspects}
\label{sec:relatedwork}
%\todoALLIn{TR 38.901, Older versions of QuaDRiGa, Other papers addressing evaluation model applied  for position. Focus on deficiencies like TRP-spatial consistency, temporal blocage and EC effects. move the QD figurev(fig.2) to SoA and adjust text in the two sections accordingly?}

\noindent
In recent years, several extensions to the  3GPP channel model~\cite{tr38901} have been introduced. Most notable for positioning, are the extensions absolute time of arrival (ATOA) model and spatial consistency. 
%This is not a problem for OFDM-based communication systems, where a UE needs frame and symbol synchronization in the time domain and therefore the absolute propagation time skew is irrelevant.
In addition to the fact that the OLOS effect is not well covered in the available models~\cite{ECpaper}, there remain 3 key challenges despite these extensions to yield more realistic positioning channel models that we describe in the following.
% ok keine bewegungsinfo

First, The simulation is initialized and not updated (drop-based) with random UE positions for which correlated path delays and departure and arrival angles are generated. Thus, the channel model does not support \textit{realistic mobility} of the communication nodes or the scattering clusters.

% hier ist ein bruch!!
% wieso gehts jetzt wieder um sc?
% gerade gings um bewegung und tracking weil alle extension nicht in der lage sind positionen zu ermöglichen die bewegung / zeit berücksichgiten 
%the spatial consistency model (SCM)~\cite{??} introduced in 3GPP 38.901 Rel. 16, enables realistic correlation in the small-scale-fading (SSF) of the channel, i.e., the positions of individual scattering clusters are now correlated for closely spaced UE locations.
Second, the recently introduced spatial consistency~\cite{tr38901} is only defined for the angles and delays of a MPC, but not for the scatterer positions themselves. As a result, small changes in the UE position may result in large changes in scatterer positions when the x,y,z-coordinates of the scatterers are calculated from the angles and delays.
Tracking the changes over time, the scatterers may move at unrealistically high velocities, that cause huge Doppler shifts.
Another consequence of this is that the propagation environment is a function of the UE position. 
This is also unrealistic for positioning applications, as they generally consider a propagation environment (e.g., buildings, cars) that are independent of the UE location. As such, machine learning models trained on such data may not accurately represent the surroundings~\cite{feigl2021ml4toa}. Spatial consistency can only be applied within the same simulation run (independent of the geometry), which can lead to an over-fitted model. 
Hence, to employ spatial consistency we introduce deterministic clusters that remain independent of the UE position. 

Third, despite the fact that the statistical properties may be dependent on deployment scenarios and environment characteristics,  the ATOA model parameters are identical for all supported NLOS scenarios. The resulting median value of the ToA error is 31~\si{ns} (approx. 9.3~\si{\m}). Given that the additional delay in the ATOA model is randomly generated, its usability for machine learning approaches is questionable.

TR38.901~\cite{tr38901} proposes an alternative approach based on a hybrid model that combines a ray-tracing model with a statistical model. However, in many cases this is not desirable due to the dependency on digitized maps for each scenario. Furthermore, ray-tracing suffers from two disadvantages:
First, for many objects, surface properties are unknown or approximated.
This affects the reflection factor and the strength of the resulting path.
Second, it is difficult to account for all static and dynamic objects in an environment to generate representative channel impulse responses (CIRs).

 For the evaluation, fusion or comparison of positioning technologies it is sufficient to generate models of a deployment that include elements with representing characteristics of a propagation effects relevant for positioning applications. This may simplify the definition of reference models and maintains the advantages of statistical models.The modelling of the attenuation or impairment of the first arriving path is essential. Especially objects close to the receiver or transmitter may cause MPCs arriving with a low delay or may obstruct the LOS path, but due to effects like diffraction a signal with minor delay relative to the LOS distance may be still detectable (OLOS reception conditions). For models such as TR38.901~\cite{tr38901} the LOS path and the ground reflection are only modeled deterministically. All other clusters are randomly generated. These other paths are only locally correlated using the spatial consistency methods such as defined by TR38.901. Spatial consistency ensures that the CIR has similar properties for UEs that are close to each other. However, several TRPs are typically required for positioning. The relationship of CIRs to different TRPs is not considered. Thus, methods that also exploit other MPCs of the CIR, such as reflections from walls, cannot be evaluated with such models.

\section{Concept of Semi-Deterministic Clusters}
\label{sec:sdc_concept}
%\todoALLIn{Align with SoA QuaDRiGa description. Decribe how the SDCs complinet an InF, UMi.. scenario Elaborate on how to introduce the SDC: single/dual/multi-bounce model; Update continuously the position of clusters}

\noindent
Positioning methods profit from the consistency and representativeness of the propagation environment. Geometry based channel models already cover this for LOS based methods. For NLOS conditions a consistent modelling of the MPCs is essential. Furthermore, for practical applications moving objects or other changes in the propagation conditions may cause variations of the CIR. Accordingly a CIR shall be considered as a composite of a static part and a time variant part. Positioning methods based on machine learning (including fingerprinting~\cite{niitsoo2018convolutional}) be suffer from these effects.    
Inline with the hybrid model we address these challenges by proposing a combination of a geometric-stochastic model such as the 3GPP-NR models and a simplified deterministic model. The deterministic model can optionally be derived from ray-tracing extracted information. Scattering positions can be obtained from geometric considerations, and the reflection factors can either be calculated from theory or matched to measurements. 

\subsection{Semi-Deterministic Clusters (SDCs)}

SDCs are clusters that have a specific position relative to a defined reference position and power relative to a defined reference power. Unlike in the classical 3GPP-NR channel model~\cite{tr38901}, SDCs are not defined by their angles or delays, but by their $x,y,z$-coordinates. In the most simple case, the SDC position is given in absolute coordinates and the power is relative to the free-space path loss (FSPL). SDCs may be generated from a known geometry, e.g., from a ray-tracing simulation, such as ground-, wall- or ceiling-reflection in indoor scenarios~\cite{JAEVTC2017}. However, other constellations are possible, e.g., positions can be relative to the TX or RX such as the reflections from the roof of a car. This allows to evaluate the impact of specific predictable signal components on the positioning performance. Typically, the SDC location and power must be obtained for the specific scenario that is evaluated. In contrast to ray-tracing-based methods, our geometry-based statistical channel model may account for moving scattering objects and offers other advantages when we generate CIRs with four types of clusters:

\begin{enumerate}
    \item \textbf{Random cluster position.}
    This method represents the classic statistical model~\cite{tr38901}.
    Typically, AoA, AoD and delay are sampled from statistical distributions with properties derived from measurements.
    They can be converted to cluster positions if a dual bounce model is adopted~\cite{JAE2017}.
    Conversely, we can randomize the cluster positions and determine AoA, AoD, and path delays.
    
    \item \textbf{Fixed (or shared) clusters.}
    The position of the cluster represents objects in the environment (importable from real measurements or ray-tracing or defined manually according a reference scenario).
    The position is the same for all links and remains fixed for the complete simulation. AoA, AoD, and delay can be calculated according to the position of the cluster, UE, and TRPs.
    
    \item \textbf{Fixed (or shared) reflector positions.}
    In case of specular reflections the effective reflection point depends on the position of the UE, TRPs, and the reflector position and orientation. Assuming the departure angle for a specular reflection is identical to the angle of arrival the effective position of the reflection point can be calculated and will be updated according to the movement of the moving device. 
        
    \item \textbf{Relative clusters.}
    The effective position of the reflection point depends on the relation of the scattering object to the position of the UE and the TRPs and is calculated according to the geometric considerations, e.g., the ground reflection changes its location based on the UE position. 
\end{enumerate}
Other parameters of the path such as signal strength or phase may be selected randomly or calculated deterministically.

\subsection{Implementation using QuaDRiGa}
%% Stephan 

% QuaDRiGa is an implementation of the models defined by 38.901 and Winner
% "Scatting cluster" 
% assuming dual bounce model the AoD, AoA and delay defines the cluster positions 
% AoD, AoA and delay statistical properties are define by Winner/38.901 

% QuaDRiGa calculates the FBS and LBS
% „fast fading effects results from change of distance to the FBS and LBS
% This generates spatial consistent fast fading effects for moving devices 
% ..... further models for spatial consistency of small scale fading
%
% \textsc{drifting} concept
%    — for drop based simulation a FBS and LBS set is calculated per drop
%    — for moving devices a sequence of „drops“ (called track) is generated. 
% The rate for update of FBS and LBS position can be selected.

% - SDC 
% -- direct definition of cluster positions
% -- statistical parameter for SDC position can be defined inline with "usage scenario" (room size, corridor parameter, selection of "important/critical scatterer")

% - deterministic scatter can be defined manually or by import of data from other tools (e.g. Ray-tracing)

% - allows definition of a reference usage scenario with surrounding objects 
%    -- "InF scenario as closed all with random position 
%    -- minor impact to "cluster based 38.901 concept"

% 

The QuaDRiGa channel model~\cite{QDWEBSITE,JAE2017,winner} provides a concept called \textsc{drifting} for the modelling of moving devices. This updates the angles and delays depending on the UE position. The change of the delay implies also a change of the phase. This is equivalent to the modelling of "Doppler effects" for moving devices. 
QuaDRiGa implements the time variant (i.e., position variant) behaviour of the channel in two steps. For a moving device the trajectory can be split in segments. Each segment is defined by a starting point of the segment and a (short) track defining the movement. For each starting point of a segment QuaDRiGa generates a random channel inline with the TR38.901 model, including spatial consistency~\cite{JAE2019}. \textsc{Drifting} calculates the change of the CIR according the UE (and/or TRP) movement. For pure drop-based simulations the segment starting points may be selected randomly and the track length may be zero. For other simulations an individual track can be assigned to each segment starting point or the segments form a continuous track. For continuous tracks the segments may be "merged" using a "cross-fade"-procedure.  

\begin{figure}[t!]
    \centering
    \includegraphics[width=1\columnwidth] {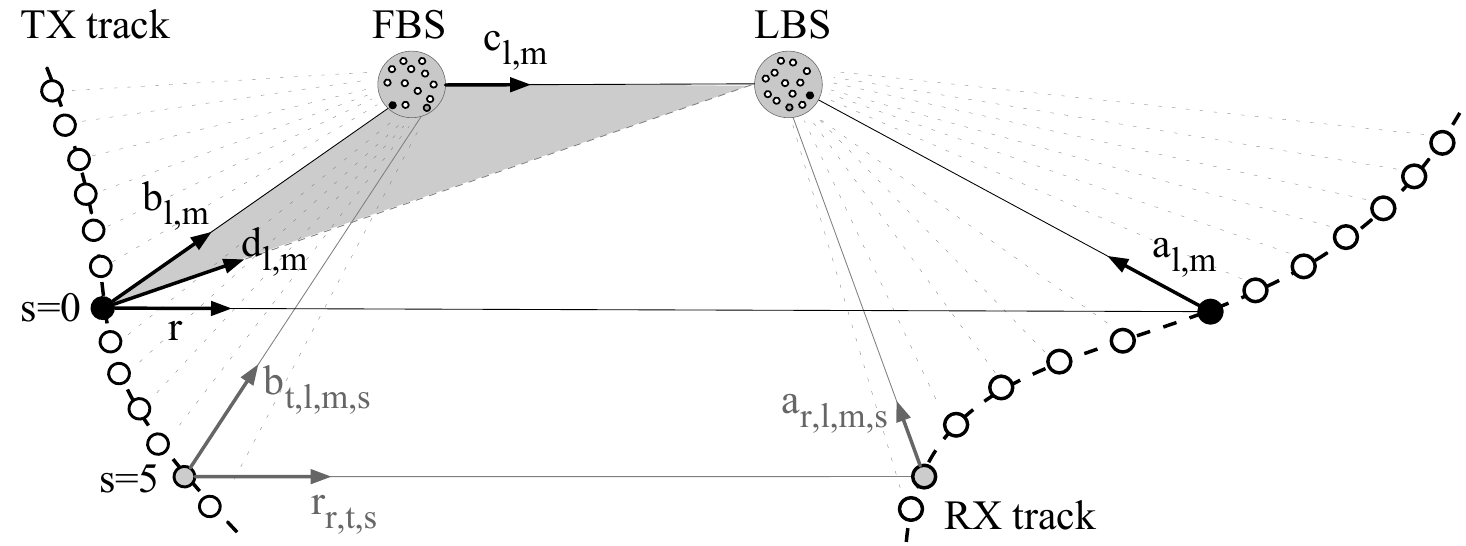}
    \caption{Illustration of the calculation of the scatterer positions and updates of the departure and arrival angles in the multi-bounce model~\cite{QDWEBSITE}.}
    \label{fig:DualBounce}
\end{figure}

The core idea is to first generate a random propagation environment and add SDCs to these (random) clusters. QuaDRiGa converts the randomly generated AoA, AoD, and delays to cluster positions assuming a dual bounce model. The SDC positions are calculated using different methods according to the SDC types defined above. To update the angles and delays for changing UE and BS (or TRP) positions \textsc{drifting} is applied. Fig.~\ref{fig:DualBounce} shows the concept. Each communication link consists of an initial transmit (TX) and receive (RX) position. A transmitted signal is reflected and scattered by objects in the environment, causing multiple copies of that signal to be received by RX. Each signal path consists of a departure direction at TX, a first-bounce scatterer (FBS), a last-bounce scatterer (LBS), and an arrival direction at RX. Departure and arrival directions are given in geographic coordinates (azimuth angle $\phi$ and elevation angle $\theta$). All random variables that determine the positions of the NLOS scatterers are spatially correlated, i.e., they depend on the initial positions of TX and RX. The 3D coordinates of the FBS and LBS can be calculated from these correlated random variables~\cite{JAE2017}.

In Fig.~\ref{fig:DualBounce} the vector $\mathbf{b}$ points from the position of a TX antenna element $t$ at snapshot $s$ to the FBS. In the second part, the vector $\mathbf{c}$ points from the FBS to the LBS, and in the third part, the vector $\mathbf{a}$ points from the RX position to the LBS. The subscripts $r$, $t$, $l$, $m$, and $s$ denote the receiving antenna element ($r$), the transmitting antenna element ($t$), the path and partial path index ($l$, $m$), and the snapshot number ($s$). Additionally, QuaDRiGa updates scatterer positions (FBS and LBS) in situations where the UE moves beyond the channels' decorrelation distance. This flexible mix of a purely stochastic simulation framework with added site-specific elements enables positioning applications in future communication networks to be evaluated.

%%%%%%%%%%%%%%%%%% solution %%%%%%%%%%%%%%%%%%%%%%%

% ??

%%%%%%%%%%%%%%%%%% setup %%%%%%%%%%%%%%%%%%%%%%%
\section{Measurement and Simulation Setups}
\label{sec:setup}
% - Evaluation of RT Delay statistics and CIR
%
% % ==> Observations: 
% - new parameter sets are required
% - more flexibility for PDP profile definition 
% - pure statistical model:  Issues to match statistics with characteristics "surrounding area"
%
% ==> introduce SDC concept
% - Three methods for SDC position definition 
%    -- statistical model 
%    -- definition according usage scenario 
%    -- combination of both
%
%\begin{figure*}[htbp]
%\centering
%\subfigure {\includegraphics[width=0.65\columnwidth]{figures/TID4003_power.png }}
%\subfigure {\includegraphics[width=0.65\columnwidth]{figures/fig_22_BW100MHz_channelID2.03.png}}
%\subfigure {\includegraphics[width=0.65\columnwidth]{figures/fig_22_BW100MHz_channelID2.7.png}}

%\caption{Measured signal strength: total power and power of the estimated first arriving path: measured (left), InF-LOS without GR according~\cite{tr38901} (middle), with SDC and GR (right)
%}\label{fig:PowerVsHeight}
%\end{figure*}

\noindent
To characterize the channel effects considered in Fig.~\ref{fig:SDC_types}, we first carry out a measurement campaign (Sec.~\ref{sec:measrement:campaign}) and then compare the results of the measurement campaign with our QuaDRiGa-based simulation setup (Sec.~\ref{sec:measrement:setup}).

\subsection{Measurement Campaign}
\label{sec:measrement:campaign}

\begin{figure}[tbp!]
    \centering
    \includegraphics[ width=1\columnwidth] {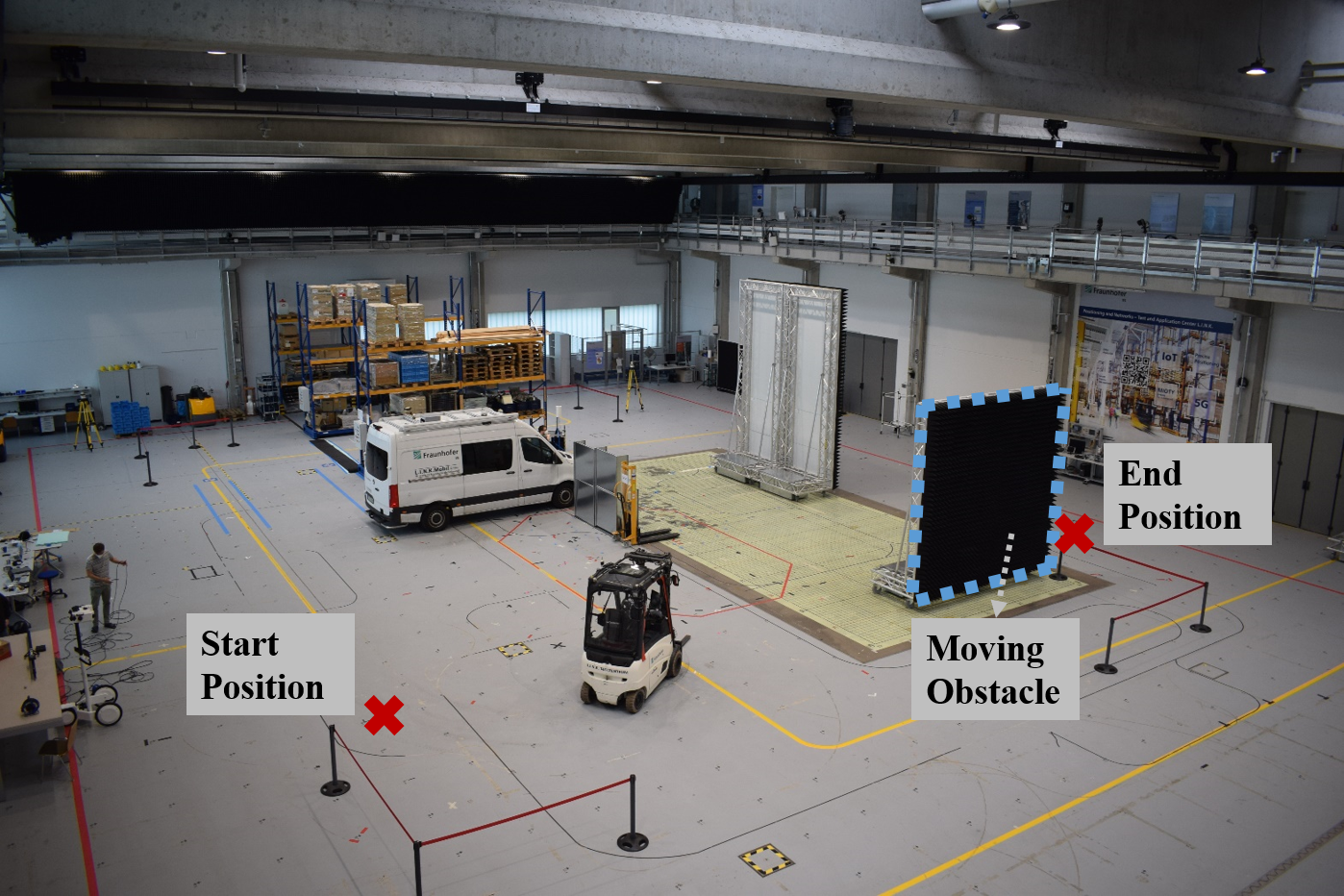}
    
    \caption{FR1-based measurement campaign. A moving obstacle (from start to end position) deterministically affects 6 base stations. The measurement setup is transferred to our simulation model to show that the latter generates comparable realistic data that represent the effects of the real environment.}
    \label{fig:MovingBlockerLINK}
\end{figure}  

The measurement setup focuses on channel effects related to signal diffraction and temporal blockage of the LOS path.
A moving obstacle temporarily blocks the LOS path between a fixed UE and the base stations.
Fig.~\ref{fig:MovingBlockerLINK} shows the single obstacle (2\,\si{m} wide and 4\,\si{m} high, with a reflective metal side and an absorbing side) and the measurement area of the test center (30\,\si{m}\,$\times$\,45\,\si{m} and a height of 10\,\si{m}).
6 base stations were mounted at different positions to provide dilution of precision for position calculation.
The environment may be viewed as a typical LOS scenario of an indoor factory.
The resulting reflective clusters are: the 4 walls, the ceiling, the floor, and the moving obstacle.
The base stations transmit a 5G-compliant positioning signal with a bandwidth of 100~\si{MHz}.
The measurements recorded by the UE were carried out at a carrier frequency of 3.75~\si{GHz}.

\subsection{Simulation Setup}
\label{sec:measrement:setup}

\begin{figure}[tp!]
    \centering
    \includegraphics[trim=10 5 5 5, clip, width=0.9\columnwidth] {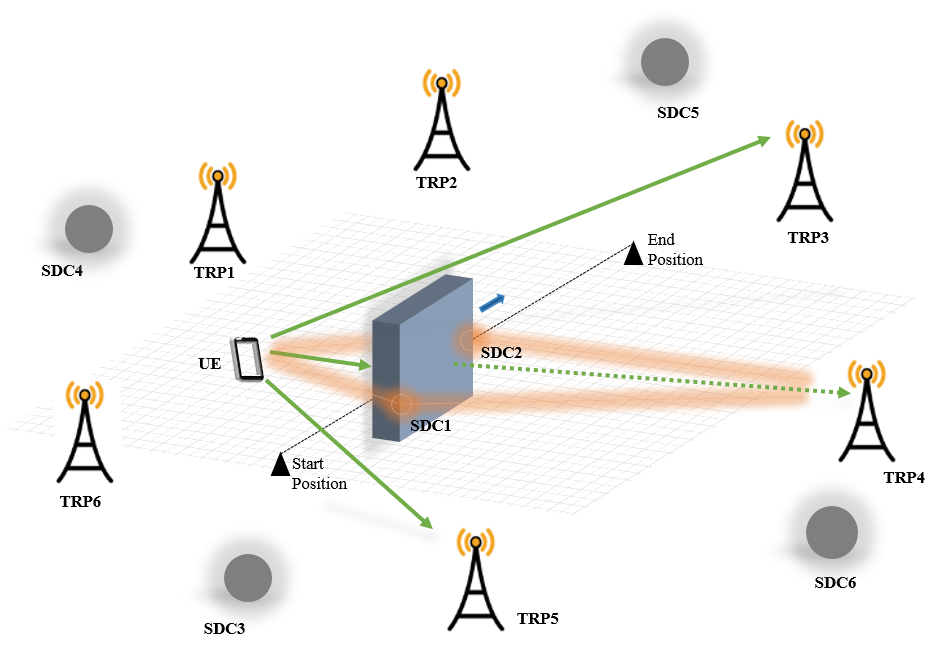}
    \caption{Simulation setup demonstrating the modeled SDCs, UE, TRPs positions, and the moving obstacle of the real environment, see Fig.~\ref{fig:MovingBlockerLINK}.}
    \label{fig:MovingAbsorber}
\end{figure}    

For the evaluation of the proposed method we recreate essential parts of the environment from the measurement campaign in the simulation by applying the SDC concept discussed in Sec.~\ref{sec:sdc_concept}.
The base stations and UE positions in the simulation and measurement setups are identical.
The channel configuration is derived from the 3GPP-InF-LOS channel parameters~\cite{tr38901}.
The same reference signal and comparable antenna configurations were used in the simulation and measurement setups.
A total of 11 complementary SDCs were used to extend the 3GPP-InF-LOS channel model for this environment. Six SDC emulate the diffraction effects at the edge of the obstacle. 5 SDCs imitate the 4 walls (South, East, North, West) and the ceiling. Ground reflection was enabled. The obstacle was modeled to block the incident path and the ground reflection. 
The ground reflection function is in line with TR38.901~\cite{tr38901}.
The effective position of the reflection point for all SDCs depends on the TRP, UE, and SDC positions.

\begin{figure}[bp!]
    \centering
    \subfigure {\includegraphics[width=0.8\columnwidth,trim=15 5 20 5, clip]{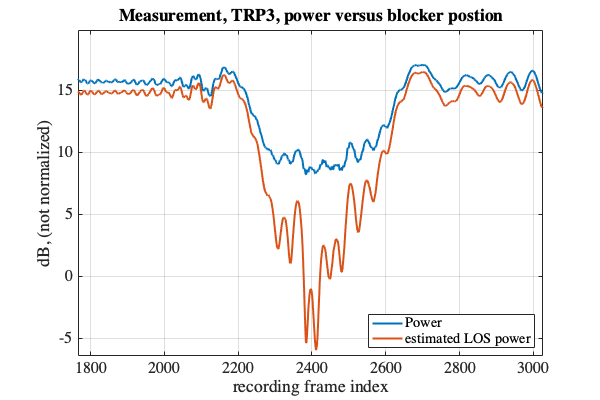}}\\
    \subfigure {\includegraphics[width=0.8\columnwidth,trim=15 5 20 5, clip]{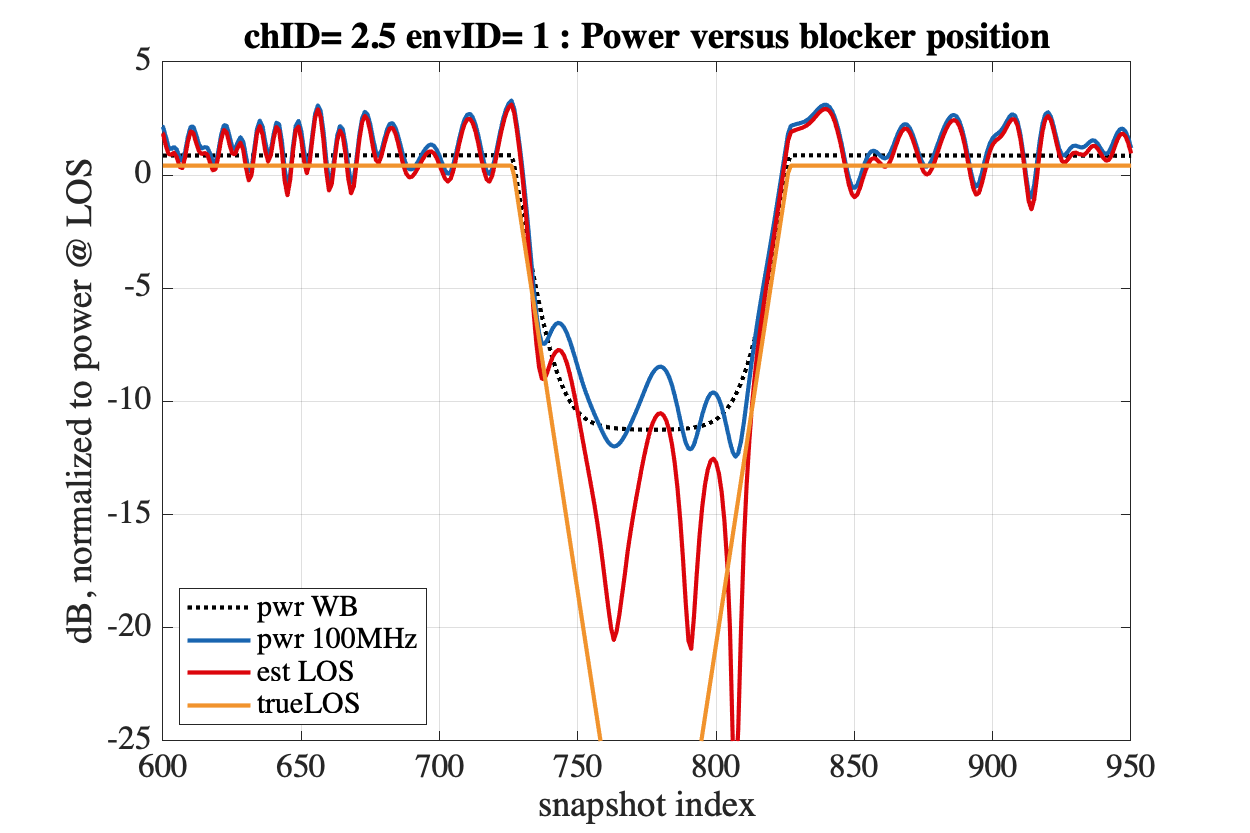}}
    \caption{Signal power of a diffraction experiment  for the moving obstacle (top: measurement; bottom: simulation using SDC).}
    \label{fig:MovingBlockerExperiment}
\end{figure}

 % \section{Characteristics of 38.901 model}
 % this is covered by the sections above

%%%%%%%%%%%%%%%%%% evaluation %%%%%%%%%%%%%%%%%%%%%%%
\section{Evaluation}
\label{sec:evaluation}
% some results showing how the observation measurement are covered by the model

\noindent
We evaluate the implementation of the SDCs for OLOS, blockage, and geometric consistency w.r.t. environment. A simulation according to the baseline models in TR38.901~\cite{tr38901} is not associated with the environment. As a result, comparing the simulations without SDC with the measurements is valueless. For the evaluations, we compare the SDC enabled simulation with the corresponding measurement occasions.

% \todoALLIn{
% - show blockage effect demonstrated with the moving blocker in figure 5
% - Discuss the CIRs
% - demonstrate how sdc can solve the ATOA in the absence of LOS
% }

\subsection{Blockage and Diffraction Observations} 
\label{sec:evaluation:obs}

Since the UE and base stations do not move, we are interested in characterizing the blockage effect from the moving obstacle on the channel. Fig.~\ref{fig:MovingBlockerExperiment} shows the measured received power between TRP3 and the UE (top) and the corresponding simulation result (bottom). We measured both the total power and the power of the first arriving path (FAP) which defines the detectable LOS path in this experiment. The measurement was carried out using time frames and we simulated using the position index. In both graphs measurement (top) and simulation (bottom), we can clearly differentiate between the true LOS power and estimated power from the FAP. 
In the measurement, between the time frames 2250 and 2600, the estimated LOS path strength is reduced by up to 20~\si{dB}. We observer similar patterns in the SDC-enabled simulation on the equivalent $x$-axis interval between position indices 700 and 800. The main power contribution in this interval is from the MPCs, showing an attenuation of around 6~\si{dB}. This "knife-edge diffraction" effect shows that our proposed complementary SDCs can accuracy model this behavior. Some fine-tuning of the parameters may be required to get a full match of measurement and simulation. But as pointed out above many parameters influence the signal strength. An optimal mapping requires the characterization of all material properties (e.g., reflection coefficients, etc.).
%All material properties such as reflection coefficient etc. must be characterized for complete match. 

We also observe fluctuations in FAP performance before and after this interval. This may result from constructive or destructive contributions of early clusters in the bandwidth-limited signal. Fig.~\ref{fig:Blockerexperiment} demonstrates the model behavior in detail, showing the generated CIR and the resulting correlation of the bandwidth-limited signal. The configured moving SDCs, indicated as "diffraction SDC" in the graphs, represent the edges of the obstacle, modelling this behavior. Other SDCs representing the walls are labeled with "Room SDC" or "GR" (ground reflection). Clusters at random positions are marked as "random MPs" (random MPCs).

Note that the constant power offset difference between measurement and simulation is due to different normalization. In the simulation, we select the magnitude of the CIR as scale. In the measurement we used the correlator output including processing gain and amplifier gains of the receiver.

%% Bild Take05_USRP_Toa_fig2_zoom3.png aus Folder "FR1_experiment/2021_05_26
%\begin{figure}[htbp]
%\centering
%\includegraphics[width=1.0\columnwidth] {figures/MovingBlockerToA.png}
%\caption{Measured ToA for moving absorber experiment}\label{fig:MovingBlockerToA}
%\end{figure}     

\subsection{Mapping of Multipath Clusters on the modelled SDCs} 
\label{sec:evaluation:sc}

\begin{figure*}[htbp]
    \centering
    \includegraphics[width=1\linewidth]{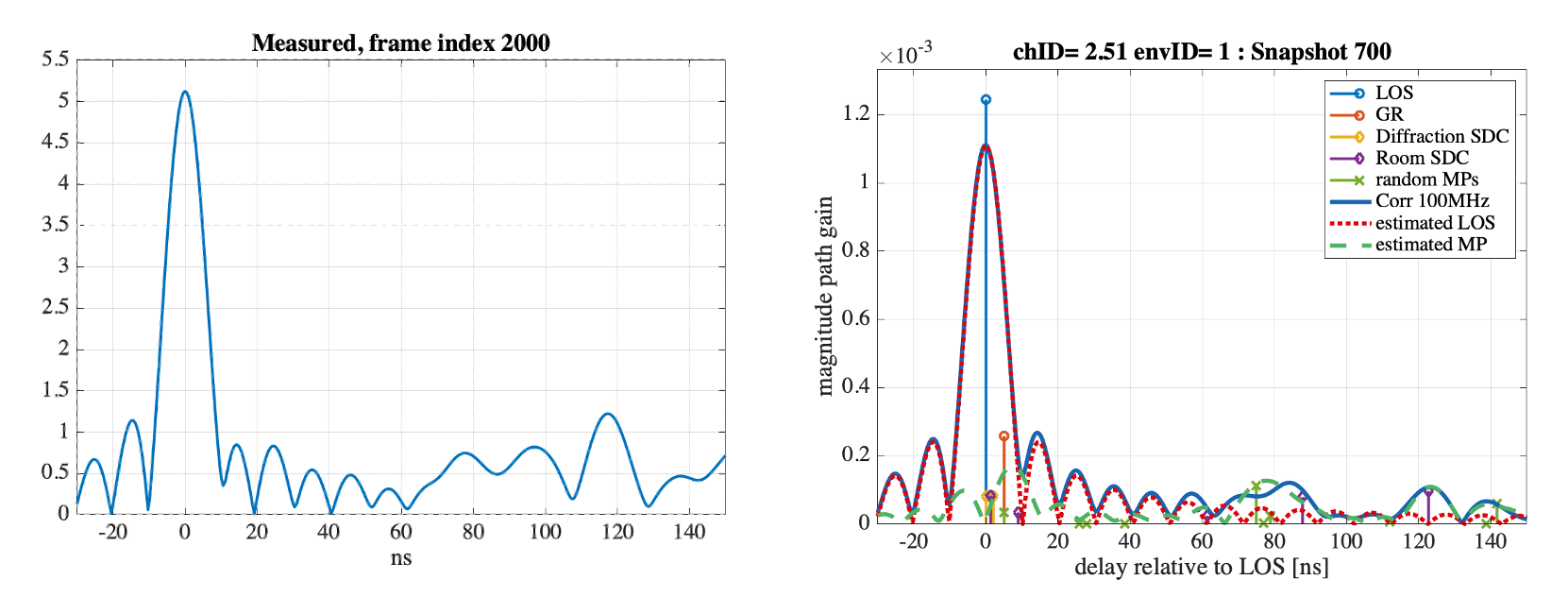}\\
    \includegraphics[width=1\linewidth]{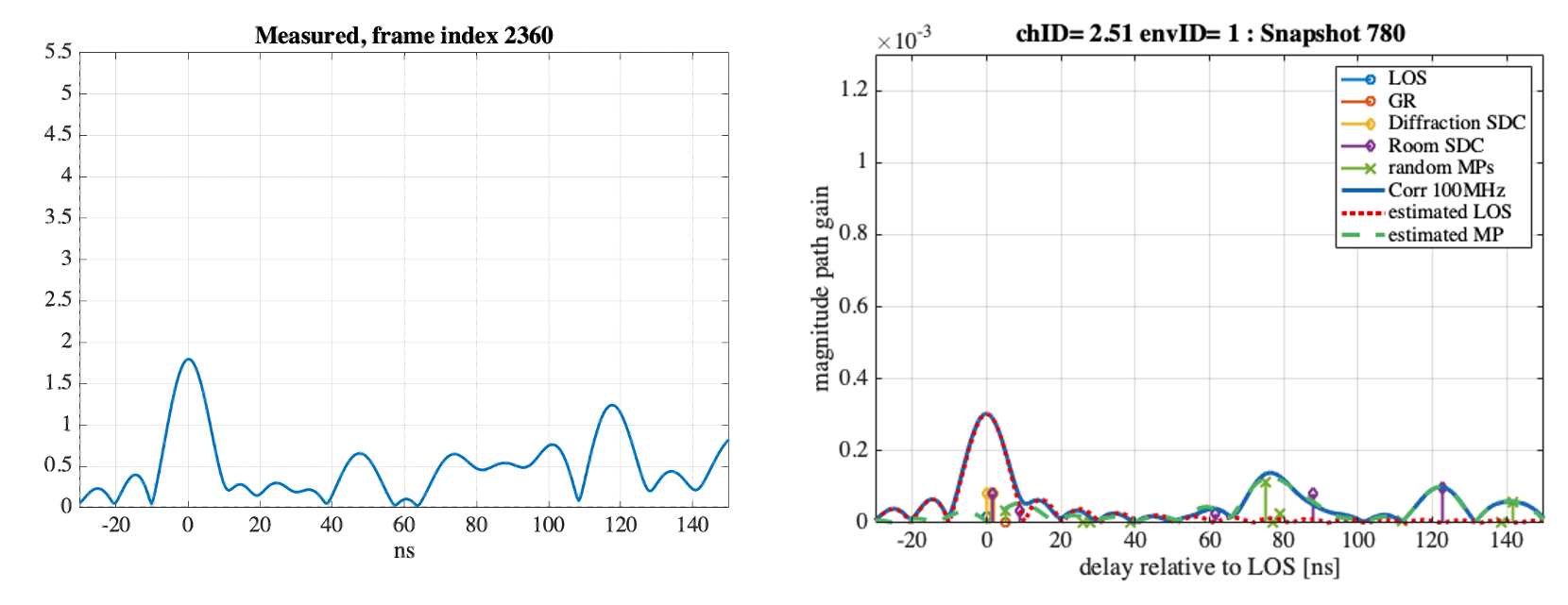}
    
    %\subfigure {\includegraphics[width=1\columnwidth]{figures/TID4003_fig14_scriptIPIN_V3.png }}
    %\subfigure {\includegraphics[width=1\columnwidth]{figures/TID4003_fig18_scriptIPIN_V3.png}}
    %\caption{Measurement at a certain positions }\label{fig:meas2} 
    
    %\subfigure {\includegraphics[width=1\columnwidth]{figures/fig_6_BW100MHz_channelID2.01.png}}
    %\subfigure {\includegraphics[width=1\columnwidth]{figures/fig_8_BW100MHz_channelID2.01.png}}
    %\caption{Simulation with SDC disabled  }\label{fig:sim1}
    
    %\subfigure {\includegraphics[width=1\columnwidth]{figures/fig_6_BW100MHz_channelID2.7.png}}
    %\subfigure {\includegraphics[width=1\columnwidth]{figures/fig_8_BW100MHz_channelID2.7.png}}
    %\caption{Simulation with SDC enabled  }\label{fig:sim2}
    
    \caption{Comparison of two (top: LOS; bottom: OLOS) exemplary measurements (left) and corresponding simulations with our SDC extension (right). In contrast to the state-of-the-art without SDC, both graphs show very similar patterns even without fine-tuning the parameters of our SDC extension.}
    \label{fig:Blockerexperiment}
\end{figure*}

Fig.~\ref{fig:Blockerexperiment} shows two sets of correlation magnitude plots at different snapshots w.r.t. TRP3-UE. For each measurement (left graph), we provide the associated simulation snapshot (right graph) and derive an estimate for the FAP and the MPCs from the measured correlation. For the analysis, we focus on measured MPCs and compare them to the simulation. 

%\red{In the first row, the equivalent simulation without SDC shows a similar contribution from the SDC enabled correlation to the measurement. "verstehe ich nicht :) ich dachte wir zeigen 1) measurements 2) sim-sdc und 3) sim+sdc?"}
%\red{Frame index 2000, as shown Fig.~\ref{fig:MovingBlockerExperiment} maps to a LOS condition. "versteh den zusammenhang nicht, wie passt das hier her?"} 

%\blue{!! Ich dachte, wir zeigen 1) Messungen 2) "sim ohne sdc" und 3) "sim mit sdc"? wo ist "sim ohne sdc" geblieben? oder sehen wir den Mehrwert von "sim mit sdc" irgendwo versteckt und beschreiben ihn einfach nicht?!!}

The top row of Fig.~\ref{fig:Blockerexperiment} shows a propagation environment with a clear LOS component that can also be seen in Fig.~\ref{fig:MovingBlockerExperiment} at measurement frame index 2000 (or simulation index 615). Different multipath sources arrive between 10~\si{ns} and 40~\si{ns} after the FAP, resulting from the moving obstacle. Additionally, later multipath delays (beyond 60~\si{ns}) are observable in the measurement and simulation associated with the environment. It is obvious that a positioning model such as machine learning may exploit such additional information to return a more accurate position even in LOS situations%Note that the plots are derived from first experiments without precise recording of the blocker position. Repetition of the experiments are subject of on-going experiments. 

The bottom row of Fig.~\ref{fig:Blockerexperiment} shows that the power for the first path is remarkably attenuated for the measurement (left) and that several MPCs beyond 40~\si{ns} show similar power. The TRP3-UE at frame index 2360 is in OLOS condition (see also snapshot index 726 Fig.~\ref{fig:MovingBlockerExperiment}). The FAP of the correlation mainly results from the "diffraction SDCs". Around 80~\si{ns} we observe "Room SDC" and a "random MP" cluster. In the bandwidth limited signal the correlation peaks overlap and may add constructive or destructive depending on the phase of the signals. The phase (and the phase difference between several components) may change if the devices move by a few millimeters. This is known as "fast fading effects" in the CIR characteristics. In the simulation we calculate the phase for the specular SDCs according the distance. For the sub-paths related to the random MPs we choose a random phase as proposed in TR38.901~\cite{tr38901}. In the SDC-enabled simulation, the SDCs contributions are consistent with the measurements. Note that even the MPCs from the modeled SDCs are preserved, for example at 90~\si{ns} and 120~\si{ns}. Again, a positioning model such as machine learning may employ such additional information to provide a more accurate position in OLOS situations when classical methods that do not employ the information fail.

Simulation results based on the previous TR38.901 model~\cite{tr38901} are not sufficient for realistic positioning evaluations. In contrast, our complementary SDCs combined with the stochastic behavior of the model describe significantly more realistic positioning properties. Our SDCs provide a geometric interpretation for the model, which is essential for evaluating positioning methods that use multipath such as machine learning.
% SDC is used to re-model a measurement scenario
% Compare measurement data and simulated data 

%%%%%%%%%%%%%%%%%% conclusion %%%%%%%%%%%%%%%%%%%%%%%
\section{Conclusion}
\noindent
This paper introduces a concept to apply complementary semi-deterministic clusters (SDC) to provide realistic simulation models, especially for more accurate and robust positioning. 
We extend the 3GPP statistical channel models to include SDCs and show that we can generate realistic measurements to enable positioning solutions that also cover OLOS situations. We validated our concept through experiments in a real environment, which provided positioning performance results comparable to those obtained from simulations using the improved channel model with SDCs.
%Classic positioning methods, such as least squares optimization for trilateriation, assume LOS conditions between transmitter and receiver and so, do not position reliably in NLOS conditions.
 The SDC-based solution can simulate representative realistic propagation environments and does not require expensive data collection in real application areas. Our concept provides a publicly available evaluation and development platform for next-generation positioning systems~\cite{QDWEBSITE}.

% - "full statistical" and "deployment EXAMPLE" based setups with ray-tracing have pros and cons
% - method adds advantages of ray-tracing based method to statistical model 
% - Advantages (randomized CIRs) of statistical models are maintained   

% - core elements of 38.901 are maintained 
% - process relationship  AoA/AoD/pathdely  <=> cluster position is inverted 

% - statistical model for SDC position is FFS
% % discuss and name outlook, what other aspects are missing and need to be addressed for "positioning channel models"%
\label{sec:conclusion}

%%%%%%%%%%%%%%%%%% references %%%%%%%%%%%%%%%%%%%%%%%
\bibliographystyle{ieeetr}%
\bibliography{bibliography}%

\end{document}